\documentclass[aps,a4paper,preprint,longbibliography]{revtex4}%
\usepackage{amsfonts}
\usepackage{amsmath}
\usepackage{amssymb}
\usepackage[dvipdfmx]{graphicx}
\usepackage{setspace}
\usepackage{bm}%
\providecommand{\U}[1]{\protect\rule{.1in}{.1in}}
\begin{document}
\title{Universality in the mechanics of soft Kirigami}
\author{Yukino Kako and Ko Okumura}
\affiliation{Physics Department and Soft Matter Center, Ochanomizu University, 2-1-1
Ohtsuka, Bunkyo-ku, Tokyo 112-8610, Japan}

\begin{abstract}
Recently, simple scaling laws concerning the mechanical response and
mechanical transition of Kirigami have been revealed through agreement between
theory and experiment for kirigami made of paper [M. Isobe and K. Okumura,
Sci. Rep. 2016]. Here, we provide experimental data obtained from kirigami
made of soft elastic sheets to demonstrate good agreement with previous
theories, although a number of assumptions in the theory are violated and the
elastic modulus is three orders of magnitude smaller in the present case. This
remarkable universality in the mechanics of Kirigami, which could be useful
for applications, is reported with physical insights based on previous theories.

\end{abstract}
\maketitle

\section{Introduction}

It has been widely known that simple patterning on sheets
\cite{miura1985,xu2017origami}, frequently motivated by Origami or Kirigami
(Japanese traditional craft technique based on folding and/or cutting paper),
could change them to mechanical metamaterials
\cite{shan2015design,bertoldiexploiting} or tunable mechanical devices
\cite{bertoldi2017flexible}, exhibiting useful mechanical and functional
properties. A simple kirigami structure is obtained by introducing a number of
parallel cuts on a sheet. Such a simple structure gives the sheet a high
extensively, even for graphene sheets
\cite{GraphenKirigami2015Nature,GraphenKirigami2014PRB}. The physical
mechanism of this high stretchability was identified as the transition from
the in-plane to out-of-plane deformation, which is accompanied by a
buckling-induced rotation of each unit of the structure
\cite{isobe2016initial}. The buckling-induced rotation has been exploited for
developing solar-tracking batteries \cite{KirigamiSolarNC2015} and dynamic
shading systems \cite{yi2018developing}. Various other cut patterns have been
studied to explore versatile possibilities of the application of kirigami
\cite{rafsanjani2017buckling,hwang2018tunable}. Frictional and interfacial
properties of kirigami have been utilized for fabricating soft actuator
\cite{rafsanjani2018kirigami} and enhancing film adhesion
\cite{zhao2018kirigami}, respectively. However, high stretchability of
kirigami remains one of the important properties of kirigami and has been
exploited in various applications, which include conducting nanocomposites
\cite{Kirigami2015NatMat}, piezoelectric materials \cite{hu2018stretchable},
metallic glass \cite{chen2018highly}, thermally responsive materials
\cite{tang2017programmable}, stretchable strain sensor \cite{sun2018kirigami}
and flexible film bioprobe \cite{morikawa2018ultrastretchable}. 

Although applications have widely been studied, the basic physical
understanding of the high extensibility of kirigami is still premature. 

Here, we investigate mechanical response of a simple kirigami structure
fabricated with sheets of soft elastic foam and surprisingly find that their
characteristics can be well captured by previous theories
\cite{isobe2016initial,isobe2019discontinuity,isobe2019continuity} although a
number of assumptions in them are violated and the elastic modulus is three
orders of magnitude smaller in the present case. This unexpected universality
in the mechanics of kirigami is interpreted in the light of previous theories.

\section{Experimental\label{S2}}

\begin{figure}[h]
\includegraphics[width=\textwidth]{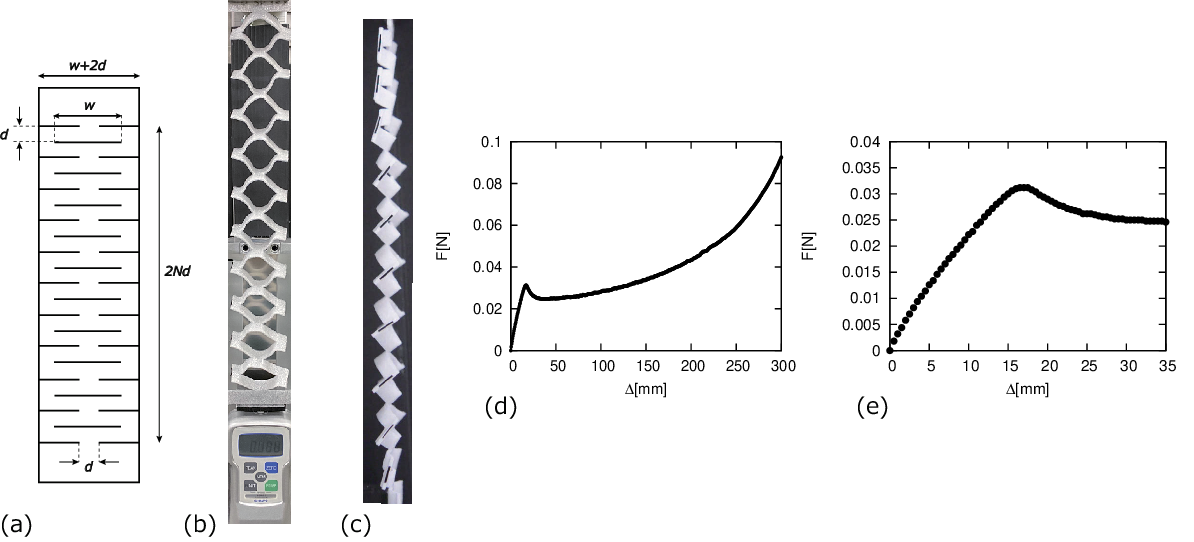}\caption{(a) Geometry of kirigami,
with the definition of kirigami parameters $d$ and $w$ (with $N=10$ throughout
this study). (b) Front view of highly extend kirigami made of soft elastic
sheet at a strain 0.825. (c) Side view at a strain 0.4. (d) Force $F$ vs
extension $\Delta$ obtained for a kirigami sample. (e) Magnified plot of the
initial regime in (d). (b) and (c) correspond to $(h,d,w)=(2,10,50)$ in mm,
while (d) and (e) to $(2.5,12,70)$.}%
\label{Fig1}%
\end{figure}

As in our previous study
\cite{isobe2016initial,isobe2019discontinuity,isobe2019continuity}, we focus
on a simple kirigami structure, characterized by three parameters, thickness
$h$ and geometrical parameters $d$, $w$, and $N$ (fixed to $N=10$, throughout
the present study), as illustrated in Fig. \ref{Fig1} (a). The structure is
composed of $N$ elemental units (width $w+2d$ and height $2d$) connected at
the top and bottom in the central area of width $d$. An elemental unit is
composed of two thin strips of height $d$, each of which will be called a half
element, connected at both lateral ends with regions of width $d$. We assign
the unit element number $n$ to each element, from $n=1$ (for the bottom unit)
to $n=N$ (for the top).

This structure is made of soft elastic sheets, commercially sold for
shock-absorbing purposes (Lightron S, Sekisui Chemical Co., Ltd.). This
material is linear elastic for small deformation up to a strain $\varepsilon
\simeq0.1$ with an elastic modulus $E\simeq$ a few MPa, which is three orders
of magnitude smaller than that of paper (fracture mechanical properties for
this material have been studied in \cite{Shiina06,Kashima2014,Takei2018}). In
the present experiment, we use two types of sheets: one with $h=2.0$ mm and
the other with $2.5$ mm, of which Young's moduli are respectively
$2.13\pm0.18$ and $1.14\pm0.01$ MPa. The structure was created using a
commercial cutting plotter (silhouette CAMEO 3, Graphtech Corp.) with a
special blade (SILH-BLADE-DEP, Graphtech Corp.). The parameters $d$ and $w$
are in the rages, 3 to 14 mm and 30 to 70 mm, respectively, where $w$ is
always at least 5 times as large as $d$. This is because previous theory is
justified in the limit $w\gg d$ and was confirmed to be practically valid when
$w$ was 5 times as large as $d$ for kirigami made of paper
\cite{isobe2016initial}.

This sample was stretched using a slider system (EZSM6D040, Oriental Motor)
with a low speed (0.5 mm/s), as in Fig. \ref{Fig1} (b). In Fig. \ref{Fig1}
(c), we see elements rotating to achieve an out-of-plane deformation, where
rotations are visible through black short lines corresponding to side edges
(height $2d$) of elemental units. Note here that the rotation angle is not
homogeneous. Even the direction of the rotation is not the same for $N$
elements! As seen in Fig. \ref{Fig1} (c), the direction in the middle area is
opposite to the one in the top and bottom area. Note that this inhomogeneity
is not reflected in the previous theories
\cite{isobe2016initial,isobe2019continuity}.

\section{Results}

\subsection{Two types of experiment and their features}

\begin{figure}[ptb]
\centering\includegraphics[width=\textwidth]{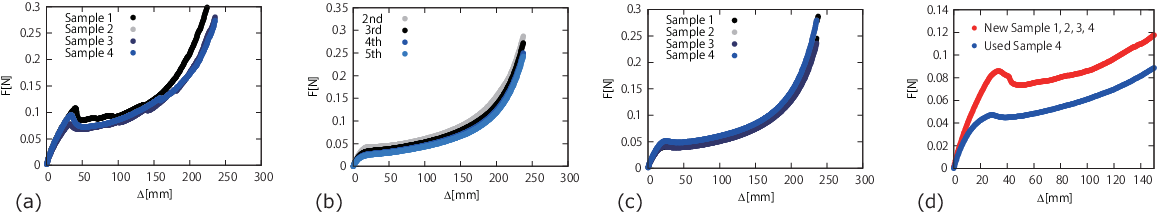} \caption{(a) $F$ vs
$\Delta$ obtained for four new samples of kirigami, showing a low
profile-reproducibility. (b) $F$ vs $\Delta$ obtained from the second to fifth
tensile experiments using an already-extended sample, with each performed
after flattening for relaxation, showing a reasonable profile-reproducibility.
(c) $F$ vs $\Delta$ obtained as the average of the second to fifth tensile
experiments from four different samples, showing a good reproducibility. (d)
$F$ vs $\Delta~$obtained by averaging results of four new-sample experiments
(performed for sample 1 to 4) and by averaging results of the second to fifth
tensile experiments performed on a single used sample, sample 4. In (a) to
(d), $(h,d,w)=(2,10,50)$ in mm.}%
\label{Fig2}%
\end{figure}

The profile of the force-extension curve obtained for new samples was not well
reproducible, as shown in Fig. \ref{Fig2} (a). This may be because of a
slightly wavy texture (the wave length and amplitude $\simeq$ cm and mm,
respectively) of the sheet surface created in the fabrication process (we
introduce the kirigami structure to sheets so that the cuts run perpendicular
to the direction of the wave vector and thus stretched kirigami in the wave
direction). However, as shown in Fig. \ref{Fig2} (b), if we reused an
already-extended sample after flattening it with a protocol specified below,
we could obtain well-reproducible curve profiles for the second to (at least)
fifth extensions. In addition, the averaged plot obtained from the used sample
was nearly independent of samples, if the parameters $h,d,$ and $w$ were the
same, as demonstrated in Fig. \ref{Fig2} (c).

The protocol for the used-sample experiment is as follows. (1) Each extension
experiment, we stretch the sample to an extension $\Delta=$ $\Delta_{b}/2$ and
keep the extension for three minutes. (2) We flatten the sample with a book
(width 17.8 cm, height 25.3 cm, and weight 452.6 g) for 15 minutes with
keeping $\Delta=0$ to restart the next stretch. In the above, $\Delta
_{b}=2Nd\left(  \sqrt{1+(w/d-1)^{2}}-1\right)  $ is a measure of extension, at
which each element starts breaking when all elements are stretched homogeneously.

Even after thus-explained flattening for relaxation, a used sample before
stretching is not flat anymore. All elements are already rotated slightly,
typically less than a few degrees, in the same direction (but with practically
negligible initial elongation on the scale of measurement). This implies that
there is no in-plane to out-of-plane transition in used samples when stretched.

Accordingly, the curve obtained from new samples and those obtained from used
samples are markedly different as shown in Fig. \ref{Fig2} (d). (In Figs.
\ref{Fig3} and \ref{Fig4} below, the results for the new-sample experiment are
given as an average of the curves obtained from 4 different new samples,
whereas those for the used-sample experiment as an average of the curves
obtained from the second to fifth extensions of a single used sample.) In the
new-sample case, the force-extension curve shows a distinct peak with a sharp
drop after the peak (although this feature tends to become less significant
after averaging), which corresponds to the in-plane to out-of-plane
transition. In contrast, in the used-sample case, the force-extension curve
shows a moderate change of slope without a sharp drop near the end of the
initial quasi-linear region.

\begin{figure}[ptb]
\centering\includegraphics[width=\textwidth]{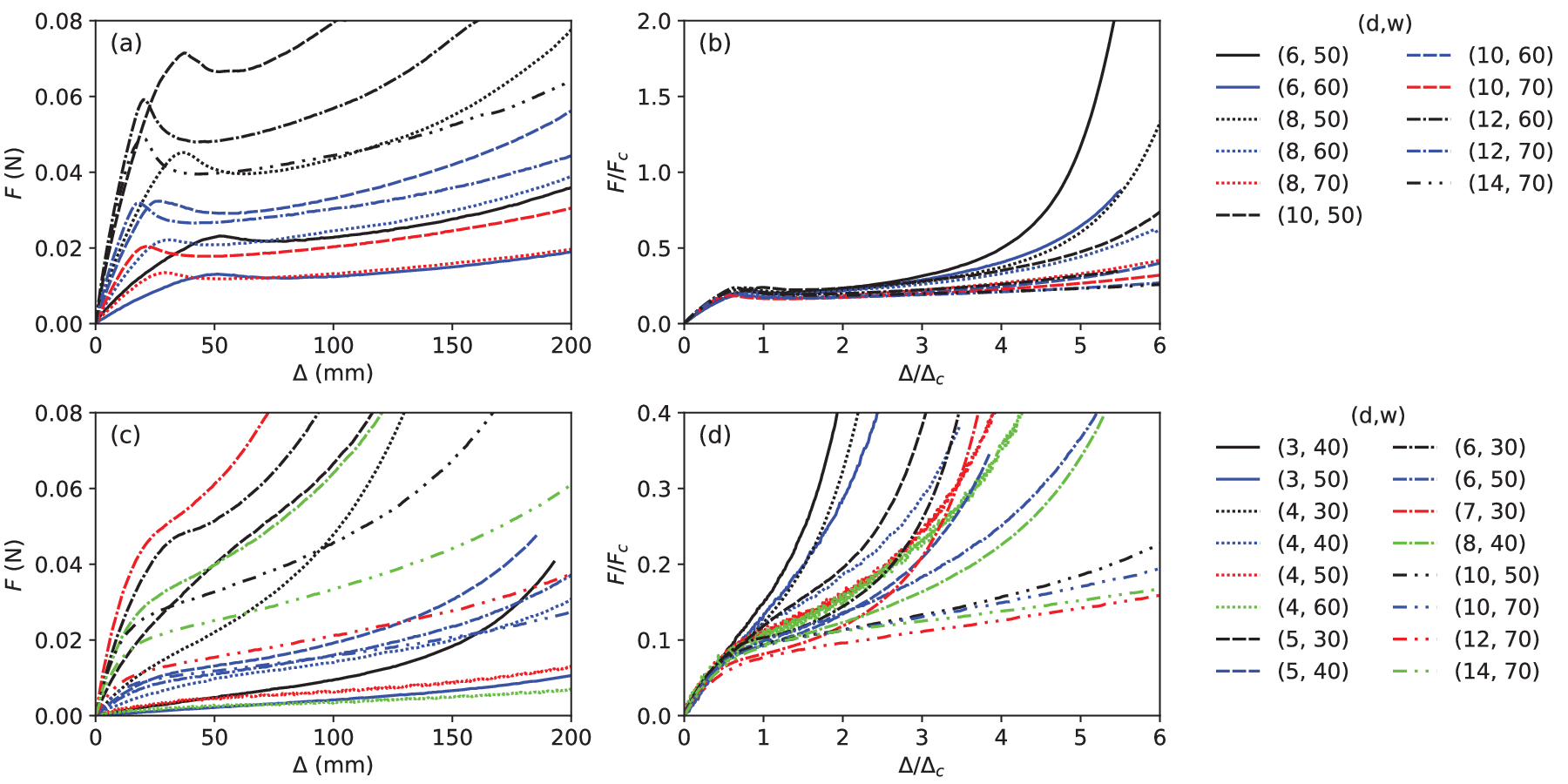} \caption{(a) $F$ vs
$\Delta$ obtained from various new samples. (b) Data in (a) plotted using
reduced axes. (c) $F$ vs $\Delta$ obtained from various used samples. (d) Data
in (c) plotted using reduced axes. Samle thickness $h$ is 2.5 mm in (a) and
2.0 mm in (c). The parameters $(d,w)$ are given in the legend in mm.}%
\label{Fig3}%
\end{figure}

\subsection{Comparison with previous scaling laws}

In recent studies, we have developed two models for the mechanics of the
extension of kirigami. The first model \cite{isobe2016initial} predicts a
discontinuity of the force-extension ($F-\Delta$) curve, somewhat similar to
the curve obtained from new samples shown in Fig. \ref{Fig2} (d). On the
contrary, the second model \cite{isobe2019continuity} predicts a continuous
$F-\Delta$, somewhat similar to the curve obtained from used samples shown in
Fig. \ref{Fig2} (d). However, both models predict the same scaling laws for
the spring constant $K$ in the initial linear regime represented by the
relation $F=K\Delta$ and the critical extension $\Delta_{c}$ at which the
mechanical transition occurs and the initial linear regime is terminated. The
scaling laws are given in the forms%
\begin{equation}
K=k/(2N)\text{ and }\Delta_{c}=2N\delta_{c}, \label{e0}%
\end{equation}
with%
\begin{equation}
k=c_{1}Eh(d/w)^{3}\text{ and }\delta_{c}=c_{2}h^{2}/d, \label{e0a}%
\end{equation}
where $c_{1}$ and $c_{2}$ are dimensionless numerical factors.

The coefficients were determined by comparing the theory and experimental data
obtained from Kent paper (whose elastic modulus is much larger than in the
present case) in \cite{isobe2016initial} as%
\begin{align}
c_{1}/(2N)  &  =0.346\pm0.006\label{c1}\\
c_{2}  &  =3.02\pm0.05 \label{c2}%
\end{align}
(We point out a small error in Fig. 3 b of \cite{isobe2016initial}: the
vertical axis is not $K_{1}/(Eh)$ but in fact $k_{1}/(Eh)$ and the line marked
'slope 3' represents $y=c_{1}x^{3}$.)

We test the relevance of these relations, in our data, shown in Fig.
\ref{Fig3}. In (a) and (c), the force-extension relation obtained from the
new-sample and used-sample experiments are respectively given for various $d$
and $w$. Note that those data were obtained with violating the condition of
homogeneous deformation of elemental units that is assumed in deriving the two
theories. In addition, in the used-sample experiment, all the elements possess
small but finite initial rotating angles. This fact is also not reflected in
previous theories and implies the absence of in-plane to out-of-plane
transition in used samples.

To demonstrate the relevance of the previously obtained scaling laws, we
renormalized the both axes $F$ and $\Delta$ by characteristic force and
extension given by $F_{c}=K\Delta_{c}$ and $\Delta_{c}$, based on Eqs.
(\ref{e0}) and (\ref{e0a}) with the coefficients given in Eqs. (\ref{c1}) and
(\ref{c2}). As a result, we surprisingly obtained clear collapses in both
cases of the new- and used-sample experiments in the initial quasi-linear
region, as shown in Fig. \ref{Fig3}(b) and (d).

\begin{figure}[ptb]
\centering\includegraphics[width=\textwidth]{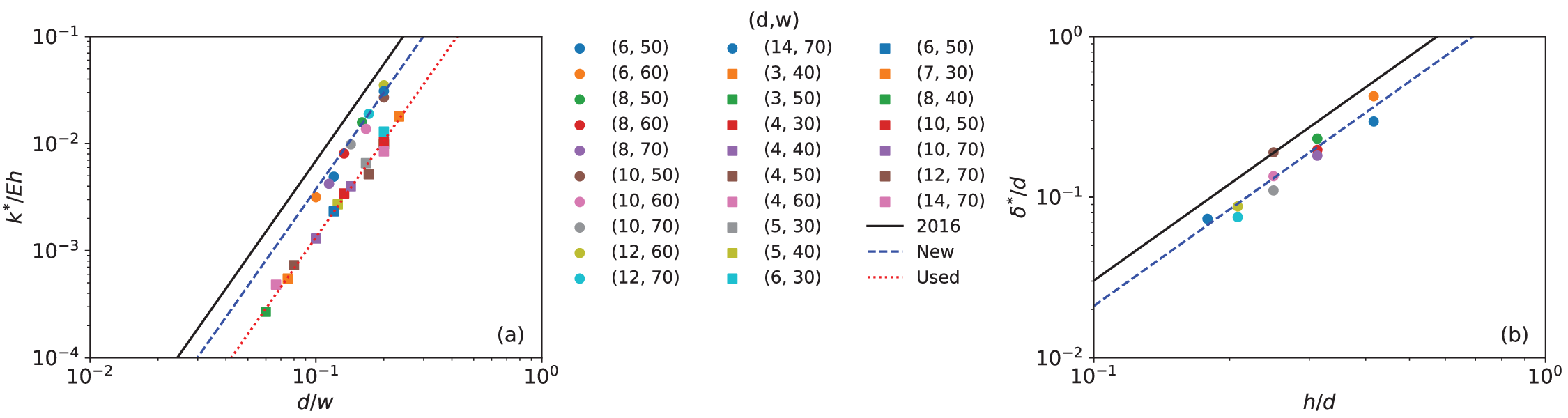} \caption{(a) Reduced
spring constant $K/(Eh)$ in the initial regime vs $d/w$. (b) Reduced critical
extension $\Delta_{c}/d$ vs $h/d$. In (a) and (b), the data are obtained from
the force-extension curve in Fig. 3 (a) or (c). The parameters $(d,w)$ are
given in the legend in mm, where circles and squares stand for the data
obtained from the new- and used-sample experiments, respectively. (In (b), the
data are shown only for the new-sample experiment. see the details for the
text). The solid line are those fitting the data in Ref.
\cite{isobe2016initial}. The dashed and dotted lines are those fitting the
data obtained in the new- and used-sample experiments, respectively.}%
\label{Fig4}%
\end{figure}

In Fig. \ref{Fig3}(b) and (d), the collapsed linear region is not terminated
at $(\Delta/\Delta_{c},F/F_{c})=(1,1)$, which implies that numerical
coefficients $c_{1}$ and $c_{2}$ are different from those given in Eqs.
(\ref{c1}) and (\ref{c2}). In addition, while the terminal points of the
linear region are located roughly at the same point $\Delta/\Delta_{c}=0.7$ in
Fig. \ref{Fig3}(b), those points are scattered in the region around
$\Delta/\Delta_{c}=0.3$ to 0.7 in Fig. \ref{Fig3}(d). This implies the
following facts. (I) In the new-sample experiment, both of the scaling laws
for $k$ and $\delta_{c}$ in Eqs. (\ref{e0}) and (\ref{e0a}) hold well with
$c_{1}$ and $c_{2}$ different from those obtained in the previous work
\cite{isobe2016initial}. (II) In the used-sample experiment, the scaling law
for $k$ is highly relevant (with a different coefficient $c_{1}$) while that
for $\delta_{c}$ is only reasonably relevant.

To quantify the above remarkable features shown in Fig. \ref{Fig3}(b) and (d),
in Fig. \ref{Fig4} (a), we evaluated the slope $K^{\ast}$ $(=k^{\ast}/2N)$ of
the initial linear regime of the $F-\Delta$ curve and plotted them on reduced
axes, represented by $k^{\ast}/Eh$ and $d/w$, because the first relation in
Eq. (\ref{e0a}) can be expressed as $k/Eh\simeq(d/w)^{3}$. As expected from a
good collapse in Fig. \ref{Fig3} (b) and (d), we found $K$ obtained from the
data in Fig. \ref{Fig3} were described by the first relation in Eq.
(\ref{e0a}) on a highly quantitative level. The difference among the data
obtained from kirigami made of paper \cite{isobe2016initial}, and those
obtained from the present new- and used-sample experiments are only numerical
factors $c_{1}/(2N)$ as indicated above, which are respectively given by Eq.
(\ref{c1})$,$ $c_{1}/(2N)=0.187\pm0.007$ and $0.0665\pm0.002$.

In Fig. \ref{Fig4} (a), as for the data obtained from the used-sample
experiments, the collapse for $k$ is distinctly clear and over a wide range.
This is because the used-sample experiment overcomes difficulty in obtaining a
good reproducibility, compared with the new-sample experiment, especially for
small $d$ and thus the used-sample experiment allows us to obtain meaningful
data for a wide range of parameters.

In Fig. \ref{Fig4} (b), the validity of the second relation in Eq. (\ref{e0a})
for the extension at the transition $\Delta_{c}$ is demonstrated for the
new-sample experiment. For this plot, we evaluated the critical extension
$\Delta^{\ast}$ $(=2N\delta^{\ast})$ from the $F-\Delta$ curve, using the
sharp peak position, in the new-sample experiment. The corresponding numerical
factor $c_{2}$ is again different: the result $c_{2}=2.31\pm0.09$ for the
present new-sample experiment differs from Eq. (\ref{c2}) obtained in
\cite{isobe2016initial}. Since the reproducibility of the new-sample
experiment is not high even after averaging, the collapse observed in Fig.
\ref{Fig4} (b) is less remarkable compared with that in Fig. \ref{Fig4} (a).

In the used-sample experiment, the in-plane to out-of-plane transition is
absent due to the initial finite rotation angle of each element. Associated
with this, the terminal points of the initial linear regime are scattered in
Fig. \ref{Fig3} (d) as discussed above. In other words, the quantity
$\Delta_{c}$ characterizes the end point of the initial quasi-linear regime,
well in the new-experiment and reasonably well in the used-sample experiment.

\begin{figure}[ptb]
\centering\includegraphics[width=\textwidth]{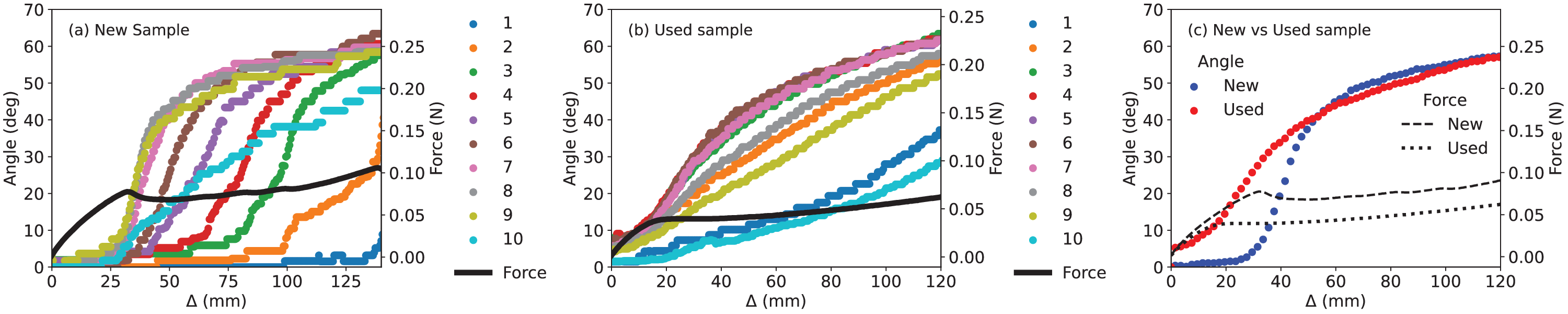} \caption{(a) Angle
$\theta$ vs extension $\Delta$ obtained for $n$th element in a single new
sample, with the $F-\Delta$ plot. (b) Corresponding plot obtained in the used
experiment for a single extension. (a) and (b) are obtained during the first
extension of sample 3 in Fig. 2 (new sample with no averaging) and at the
second extension of sample 1 in Fig.2 (used sample). The set $(h,d,w)$ is
$(2,10,50)$ in mm. (c) $\theta$ vs $\Delta$ shown as averages of the data for
elements for\ which rotation starts earlier in the new-sample and used-sample
experiment. See the text for the details on averaging.}%
\label{Fig5}%
\end{figure}

\subsection{Rotating angle and mechanical response}

As already mentioned, we have developed two models: one disallows the
coexistence of the in-plane and out-of-plane deformation and predicts
discontinuous drop of force as a function of elongation at the transition
point \cite{isobe2016initial}, and the other allows the coexistence and
predicts continuous change of force at the transition
\cite{isobe2019continuity}. In the latter work \cite{isobe2019continuity}, we
further consider the energy of kirigami as a function of the rotation angle,
in which the rotation angle of each unit elements are assumed to be
homogeneous: for a given elongation, all the unit elements rotate the same
angle. As a result, in the continuous model, we found a striking similarity to
the critical phenomena in thermodynamic phase transition
\cite{cardy1996scaling,goldenfeld2018lectures}: the energy as a function of
the angle possesses a single minimum at $\theta=0$ for small elongations
but\ two minima at $\theta=\pm\theta_{0}(\Delta)$ for large deformations with
$\theta_{0}(\Delta)\sim(\Delta-\Delta_{c})^{1/2}$, which means
\begin{equation}
\theta_{0}(\Delta)\sim\left\{
\begin{array}
[c]{ccc}%
0 &  & \Delta\leq\Delta_{c}\\
(\Delta-\Delta_{c})^{1/2} &  & \Delta>\Delta_{c}%
\end{array}
\right.  . \label{order}%
\end{equation}
The theoretical structure is mathematically equivalent to Landau theory of the
second-order transition, if we regard $\theta$ and $1/\Delta$ as the order
parameter (such as the magnetization) and the temperature, respectively. In
other words, the continuous model predicts a continuous change of the rotation
angle with a critical exponent $1/2$ at the transition. On the contrary, we
showed in \cite{isobe2019continuity} that the discontinuous model proposed in
\cite{isobe2016initial} exhibits a discontinuous jump of the angle at the
transition point.

Motivated by these theoretical predictions, in Fig. \ref{Fig5} (a) and (b), we
show results of the measurement of the rotation angle $\theta$ of each
elemental unit as a function of extension $\Delta$, obtained in the new-sample
and used-sample experiments, respectively, with the $F-\Delta$ relation
superposed, although the present case is different from the theory in that the
rotation angle is not the same for unit elements for a given $\Delta$. In Fig.
\ref{Fig5} (a) obtained from the new-sample experiment, corresponding to the
observation for inhomogeneity in elemental-unit rotations mentioned earlier
for Fig. \ref{Fig1} (c), as the elongation $\Delta$ increases the rotation
starts from 7th to 9th elements (the other elements start rotating at larger
values of $\Delta$). In addition, the inflection point $\Delta=\Delta_{\inf}$
(where the angle starts a sharp increase) on the $\theta-\Delta$ curves of
such elements which start rotating "earlier" than the others roughly
corresponds to the end point of the initial quasi-linear regime $(\Delta
=\Delta_{c})$ on the $F-\Delta$ curve, which is superposed on the plot.
Although elemental unit numbers for which the rotation start "earlier" (i.e.,
at a smaller $\Delta$) depend on the sample, a good correlation between the
inflection points $\Delta=\Delta_{\inf}$ of such "earlier" $\theta-\Delta$
curves and the end point $\Delta\sim\Delta_{c}$ was a universal feature,
regardless of the data obtained in the new-sample and used-sample experiments,
as we can confirm in Fig. \ref{Fig5} (b).

In Fig. \ref{Fig5} (c), we show averages of the "earlier" $\theta-\Delta$
curves obtained in the new-sample and used-sample experiments using sample 1
to 4 discussed in Fig. 2, as specified below, with the force curves. For the
new-sample experiment, we showed an average of the $\theta-\Delta$ profiles
for the three "earlier" unit elements (i.e., three unit elements whose
rotation start at the 1st to 3rd lowest values of $\Delta$) of sample 1 to 4.
For example, in Fig. 5 (a) obtained during the first extension of sample 3
(new sample), $n=7,8,9$ are such three "earlier" unit elements. In other
words, the average is taken over 12 profiles (3 profiles from each of 4
samples). For the used-sample experiment, we showed an average of the profiles
for the three "earlier" unit elements of the second to fifth stretch (i.e., 4
stretches), all obtained from used sample 3 (the average is again taken over
12 profiles).

Figure \ref{Fig5} (c) thus obtained reveals the above-mentioned universal
feature more clearly irrespective of whether the data are obtained from new or
used samples: the inflection point $\Delta=\Delta_{\inf}$ of the average of
"early" $\theta-\Delta$ curve roughly corresponds to the end point $\Delta
\sim\Delta_{c}$ of the linear regime of the $F-\Delta$ curve. In other words,
the initial quasi-linear regime corresponds to the region in which the
rotation angle are almost fixed and when rotation angle starts a sharp
increase without jump (continuous increase) the force makes a transition into
the second regime. This continuous change of the angle is consistent with the
continuous theory \cite{isobe2019continuity}, which predicts a continuous
change of the angle, rather than the discontinuous theory
\cite{isobe2016initial}, which predicts a discontinuous jump (although the
first derivative of $\theta$ with respect to $\Delta$ is continuous, which is
different from the behavior of Eq. (\ref{order}) at $\Delta=\Delta_{c}$).

However, the two averaged curves for the rotation angle obtained from new
samples and those from used samples are clearly different, in that in the
used-sample experiment the rotation angle starts from a finite value while in
the new-sample experiment the angle starts from zero and increases sharply
from a certain extension (transition point). Accordingly, the angle dependence
on the extension in Eq. (\ref{order}) obtained in the continuous theory
\cite{isobe2019continuity} is very similar to the averaged the $\theta-\Delta$
curve obtained from new samples in Fig. \ref{Fig5} (c) except in the vicinity
of the transition point, but different from the corresponding curve from used
samples in Fig. \ref{Fig5} (c) in that the rotation angle before the sharp
increase is a finite angle.

This finite angle before the sharp increase observed in the used-sample data
implies a strong tendency of coexistence of the in-plane and out-of-plane
deformations. As already suggested, in the continuous theory
\cite{isobe2019continuity} we considered a general deformation in which the
in-plane and out-of-plane deformations coexist (while such a coexistence is
forbidden in the latter) and showed that the former contributes to the linear
regime. In the used-sample experiment, the linear regime, which implies the
existence of an in-plane deformation, is observed when the rotation angle is
finite, which means the existence of a finite out-of-plane deformation: the
in-plane and out-of-plane deformations coexist at a significant level even in
the initial linear regime in the case of used samples.

\section{Conclusion}

We examined mechanical response of a kirigami made of soft elastic sheets.
Although the response was not highly reproducible for a new sample but fairly
reproducible for used samples, we successfully quantify the response by
focusing two types of averaged results: one for new samples and the other for
used samples. We confirmed the existence of the initial linear regime in the
force-elongation curve in both cases as in previous studies while the in-plane
to out-of-plane transition, which is also a feature of previous study, was
clearly observed only for the new-sample experiment. We surprisingly found
previously obtained scaling laws successfully characterize the mechanical
response although a number of assumptions in previous theory such as
homogeneity in rotation angle are violated and the elastic modulus is three
orders of magnitude smaller in the present case, revealing universality of the
mechanical response of kirigami. We further quantify the correlation between
the response of rotation angle and that of force as a function of elongation
to find the initial linear regime of the force-elongation curve corresponds to
a quasi-plateau regime of the rotation-elongation curve and the linear regime
ends when rotation starts in some of the elements. This tendency can be
qualitatively interpreted in the light of a previous theory in which the
coexistence of the in-plane and out-of-plane deformations is allowed. However,
applicability of previously obtained scaling laws in the present case at a
quantitative level is still a surprise. To understand this, we need further
theoretical consideration. For example, in the present case, the effect of
plastic deformation should be important, especially in the used-sample
experiment, and thus, a quantitative theory including the effect of plastic
deformation would be promising. We expect that the universal mechanical
features revealed in the present study motivate future theoretical development
and are useful for various applications.

\section*{Acknowledgements}

This work was partly supported by JSPS\ KAKENHI Grant Number JP19H01859.

\end{document}